# Growth and superconducting anisotropies of F-substituted $LaOBiSe_2$ single crystals

Masanori NAGAO[1,2*], Masashi TANAKA[2], Satoshi WATAUCHI[1], Isao TANAKA[1] and Yoshihiko TAKANO[2]

[1]*University of Yamanashi, 7-32 Miyamae, Kofu, 400-8511*

[2]*National Institute for Materials Science, 1-2-1 Sengen, Tsukuba, 305-0047*

**Abstract**

F-substituted $LaOBiSe_2$ single crystals were grown using CsCl flux. The obtained single crystals showed a plate-like shape with a size of about 1.0 mm square. The *c*-axis lattice constant of the grown crystals was determined to be 14.114(3) Å. The superconducting critical temperature of the single crystal was approximately 3.5 K. The superconducting anisotropies were determined to be 49 and 24 using the upper critical field and the effective mass model, respectively.

KEYWORDS: $BiSe_2$-based selenide superconductor, single crystals growth, cesium chloride

flux

*E-mail address: mnagao@yamanashi.ac.jp

*Corresponding Author

Masanori Nagao

E-mail address: mnagao@yamanashi.ac.jp

Postal address: University of Yamanashi, Center for Crystal Science and Technology

Miyamae 7-32, Kofu 400-8511, Japan

Telephone number: (+81)55-220-8610

Fax number: (+81)55-254-3035

**Main Text**

$BiS_2$-layered sulfide superconductors, such as $Bi_4O_4S_3$[1], $RO_{1-x}F_xBiS_2$ (R=La, Ce, Pr, Nd, Yb)[2-6] and $Sr_{1-y}La_yFBiS_2$[7], have been extensively studied. However, there are fewer papers about $BiSe_2$-layered selenide. Recently, A. Krzton-Maziopa *et al.* reported superconductivity in $LaO_{0.5}F_{0.5}BiSe_2$ polycrystalline sample with a superconducting transition temperature ($T_c$) of around 2.5 K[8]. $LaO_{0.5}F_{0.5}BiSe_2$ is a bismuth oxyselenide layered superconductor. Its crystal structure is similar to that of $BiS_2$-layered superconductor $LaO_{1-x}F_xBiS_2$ in which the site of S is completely substituted by Se. $LaO_{0.5}F_{0.5}BiSe_2$ has a tetragonal structure with the space group *P4/nmm*[8]. We expect that investigation of the $BiSe_2$-layered superconductors clarify the mechanism of not only $BiSe_2$- but also $BiS_2$-layered superconductors. Thus, single crystals of $BiSe_2$ layered superconductors attract much attention because it enables us to analyze the intrinsic physical properties. $RO_{1-x}F_xBiS_2$ (R=La, Ce, Nd) single crystals were grown using an alkali metal chloride flux in vacuum[9-11]. Therefore, various $RO_{1-x}F_xBiSe_2$ single crystals could reveal the further investigation of superconducting properties. Very recently, the structural analysis of single crystal has been reported[12].

In this paper, we report the successful growth of F-substituted $LaOBiSe_2$ single crystals using CsCl flux with the melting temperature about 645 °C in a quartz tube sealed in a vacuum. The composition and superconducting anisotropies were investigated to characterize the properties of the single crystals. The *c*-axis lattice constant of the grown crystals was determined by X-ray diffraction.

Single crystals of F-substituted $LaOBiSe_2$ were grown in a quartz tube sealed in vacuum by a high-temperature flux method. The following raw materials were used: La, Bi, Se, $Bi_2Se_3$, $Bi_2O_3$ and $BiF_3$. The raw materials were weighed with a nominal composition of $LaO_{0.5}F_{0.5}BiSe_2$. The raw materials of 0.8 g were mixed with CsCl flux of 5.0 g, and the mixture was sealed into a quartz tube under vacuum. The mixture was heated at 900 °C for 10 h, cooled down slowly at 1.0°C/h to 650 °C, and then was furnace-cooled to room temperature. After that, the quartz tube was opened under air atmosphere, and distilled water was added into the quartz tube. The product was filtered and washed with distilled water.

The crystal structure and morphology of the single crystals were evaluated by X-ray diffraction (XRD) analysis using Cu$K\alpha$ radiation and scanning electron microscopy (SEM), respectively. The transport properties of the single crystals were measured by the standard four-probe method with constant current density ($J$) using a Quantum Design Physical Property Measurement System (PPMS). The onset of the superconducting transition temperature ($T_c^{onset}$) was defined as the inflection point in resistivity–temperature ($\rho$-$T$) curve. The zero resistivity ($T_c^{zero}$) was determined considering the criterion of a resistivity of 0.1 μΩcm in $\rho$-$T$ characteristics. We measured the angular ($\theta$) dependence of resistivity ($\rho$) in the flux liquid state under various magnetic fields ($H$) and calculated superconducting anisotropy ($\gamma_s$) using the effective mass model[13-15].

Figure 1 shows a typical SEM image of F-substituted $LaOBiSe_2$ single crystal. The obtained single crystals showed plate-like shape of 1.0 mm square and 20–50 μm thickness. Figure 2 shows an XRD pattern of the well-developed plane of F-substituted $LaOBiSe_2$ single crystal. The presence of only 00*l* diffraction peaks of the $LaOBiSe_2$ structure indicates that *ab*-plane is well-developed. The *c*-axis lattice constant is 14.114(3) Å. As the results of electron probe microanalysis (EPMA), Cs and Cl of the flux components were not detected with a minimum sensitivity limit of 0.1 wt% in the grown single crystals whereas it was confirmed that the grown crystals contain La, Bi, Se, O and F.

Figure 3 shows the $\rho$-$T$ characteristics along the *ab*-plane of F-substituted $LaOBiSe_2$ single crystal. Superconducting transition was observed at around 3.5 K. The $T_c^{onset}$ is 3.75 K and the $T_c^{zero}$ is observed at 3.05 K. The resistivity drops at 3.4-3.7 K. Then, the resistivity is tailing to above 3.05 K and showing zero resistivity with the criterion (0.1 μΩcm) at 3.05 K. The resistivity shows a metallic behavior above 12 K, whereas the resistivity reveals a semiconducting-like behavior below 12 K. That is similar to the behavior of polycrystalline samples[8)]. Figure 4 shows the temperature dependence of the resistivity below 5 K under 0-9.0 T of magnetic field ($H$) parallel to the (a) *ab*-plane and (b) *c*-axis. The suppression of the critical temperature under the magnetic field applied parallel to the *c*-axis is more significant than that of *ab*-plane. The field dependence of the $T_c^{onset}$ under the magnetic field ($H$) parallel to the *ab*-plane ($H$//*ab*-plane) and *c*-axis ($H$//*c*-axis) is plotted in Fig. 5. The linear extrapolation of $T_c^{onset}$ with $H//ab$-plane and $H//c$-axis approach the value of 28 T and 0.57 T, respectively. The upper critical field $H^{//ab}{}_{C2}$ and $H^{//c}{}_{C2}$ are predicted less than 28 T and 0.57 T, respectively. We evaluated the superconducting anisotropy ($\gamma_s$) from ratio of the upper critical field using the following equation:

$$\gamma_s = H^{//ab}{}_{C2}/H^{//c}{}_{C2} = \xi_{ab}/\xi_c \ (\xi\text{:coherence length}), \quad (1)$$

was around 49. This result indicates that F-substituted $LaOBiSe_2$ superconductor has high anisotropy.

We performed that the superconducting anisotropy of F-substituted $LaOBiSe_2$ was estimated from another approach, which was using effective mass model[13)]. The angular ($\theta$) dependence of resistivity ($\rho$) was measured at various magnetic fields ($H$) in the flux liquid state to estimate the superconducting anisotropy ($\gamma_s$) of the grown F-substituted $LaOBiSe_2$ single crystal, as introduced in Refs. 14) and 15). Reduced field ($H_{red}$) is evaluated using the following equation for the effective mass model:

$$H_{red} = H(\sin^2\theta + \gamma_s^{-2}\cos^2\theta)^{1/2}, \quad (2)$$

where $\theta$ is the angle between the $ab$-plane and the magnetic field[13)]. $H_{red}$ is calculated from $H$ and $\theta$. Superconducting anisotropy ($\gamma_s$) was estimated by the best scaling for $\rho$-$H_{red}$ relations. Figure 6 shows the angular ($\theta$) dependence of resistivity ($\rho$) at various magnetic fields ($H$ = 0.05-9.0 T) in the flux liquid state for F-substituted $LaOBiSe_2$ single crystal. The $\rho$-$\theta$ curve was represented by a two-fold symmetry. The $\rho$-$H_{red}$ scaling obtained from the $\rho$-$\theta$ curves in Fig. 6 using Eq. (2) is shown in Fig. 7. The scaling is obtained by taking $\gamma_s = 24$, shown in Fig. 7.

These results suggested that the superconducting anisotropy of F-substituted $LaOBiSe_2$ single crystal was estimated to be 24-49. The values of $\gamma_s$ are different between Eq. (1) and Eq. (2). That reason is unclear at this moment. Further investigation is required for the origin of this differential. The value of superconducting anisotropy of F-substituted $LaOBiSe_2$ single crystal is similar to that of F-substituted $LaOBiS_2$ single crystal[11)].

We have successfully grown the F-substituted $LaOBiSe_2$ single crystals using CsCl flux. The F-substituted $LaOBiSe_2$ single crystals of about 1.0 mm square were obtained. The *c*-axis lattice constant of the grown crystals was determined to be 14.114(3) Å. The $T_c^{onset}$ and $T_c^{zero}$ of the single crystal were 3.75 and 3.05 K, respectively. The values of $\gamma_s$ were determined to be 49 and 24 using the upper critical field (Eq. (1)) and the effective mass model (Eq. (2)), respectively.

**Acknowledgment**

The authors would like to thank Dr. A. Miura of University of Yamanashi for useful discussions.

## Figure captions

FIG. 1. SEM image of F-substituted $LaOBiSe_2$ single crystal.

FIG. 2. XRD pattern in well-developed plane of F-substituted $LaOBiSe_2$ single crystal.

FIG. 3. Resistivity–temperature ($\rho$-$T$) characteristics along the *ab*-plane of F-substituted $LaOBiSe_2$ single crystal. The inset is an enlargement of the superconducting transition.

FIG. 4. (Color online) Temperature dependence of resistivity for F-substituted $LaOBiSe_2$ single crystal under 0-9.0 T of magnetic fields parallel to the (a) *ab*-plane and (b) *c*-axis.

FIG. 5. The field dependence of the $T_c^{onset}$ under the magnetic field ($H$) parallel to the *ab*-plane ($H$//*ab*-plane) and *c*-axis ($H$//*c*-axis). The lines are liner fits to the data.

FIG. 6. (Color online) Angular $\theta$ dependence of resistivity $\rho$ in flux liquid state at various magnetic fields for F-substituted $LaOBiSe_2$ single crystal.

FIG. 7. (Color online) The same set of data as Fig. 6, scaling of angular $\theta$ dependence of resistivity $\rho$ at a reduced magnetic field $H_{red} = H(\sin^2\theta+\gamma_s^{-2}\cos^2\theta)^{1/2}$.

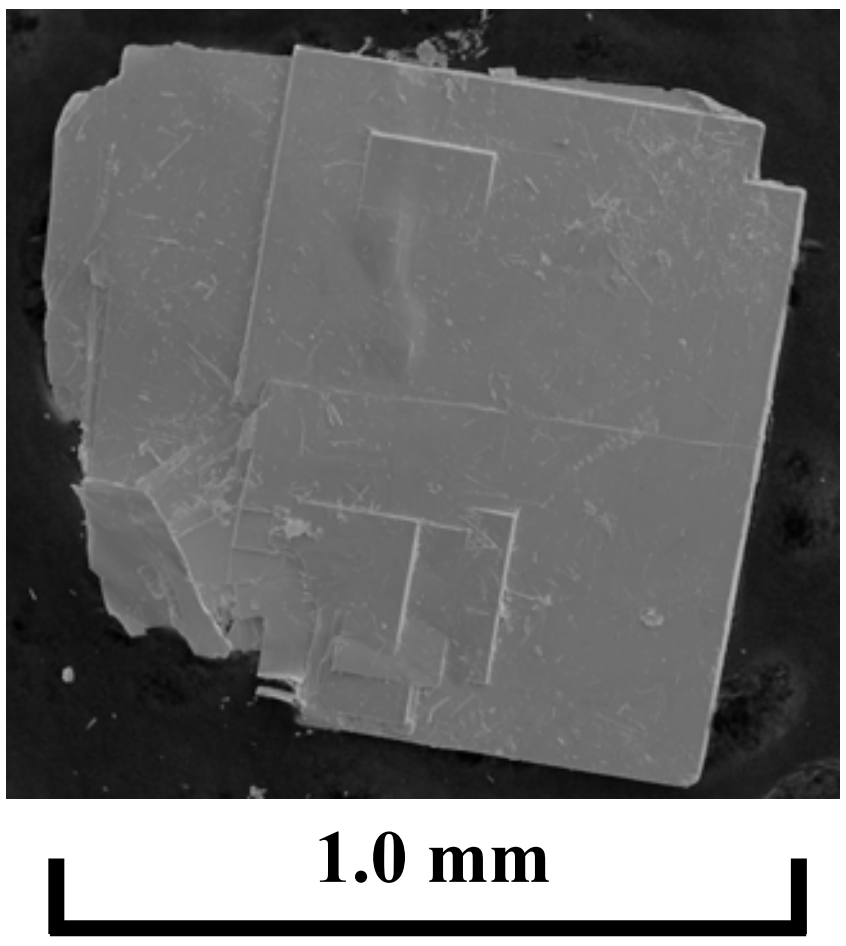

**Figure 1**

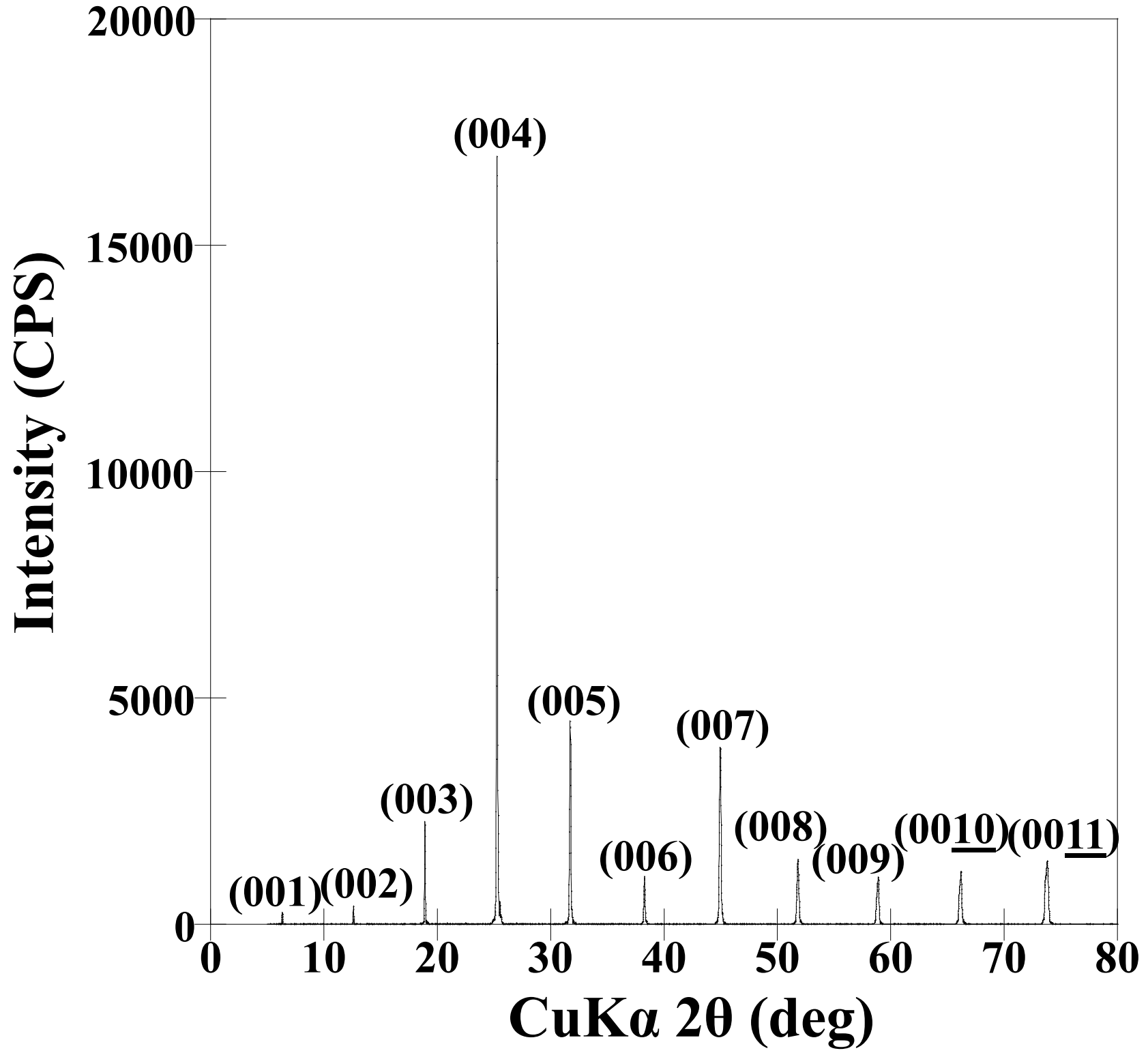

Figure 2

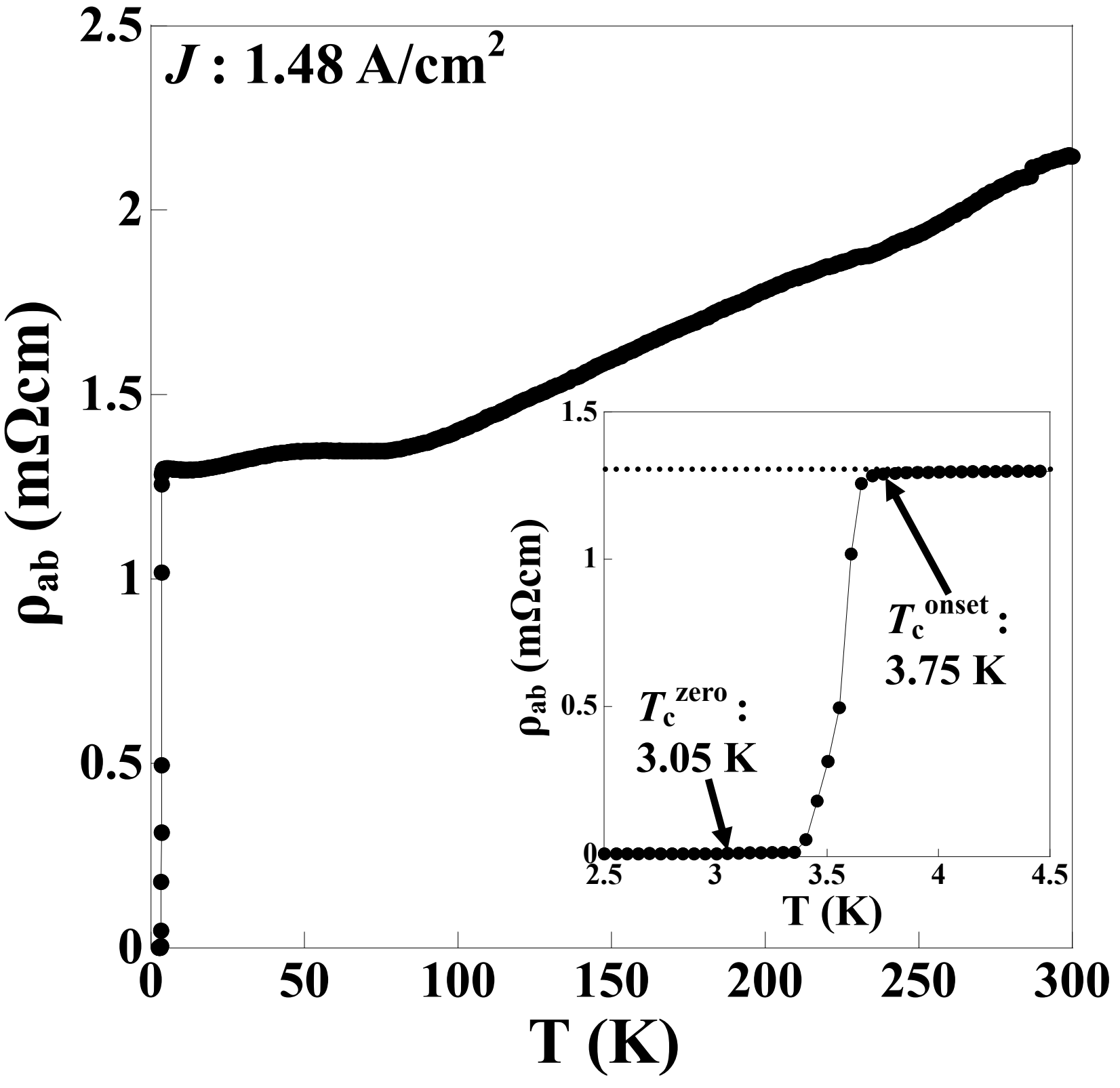

**Figure 3**

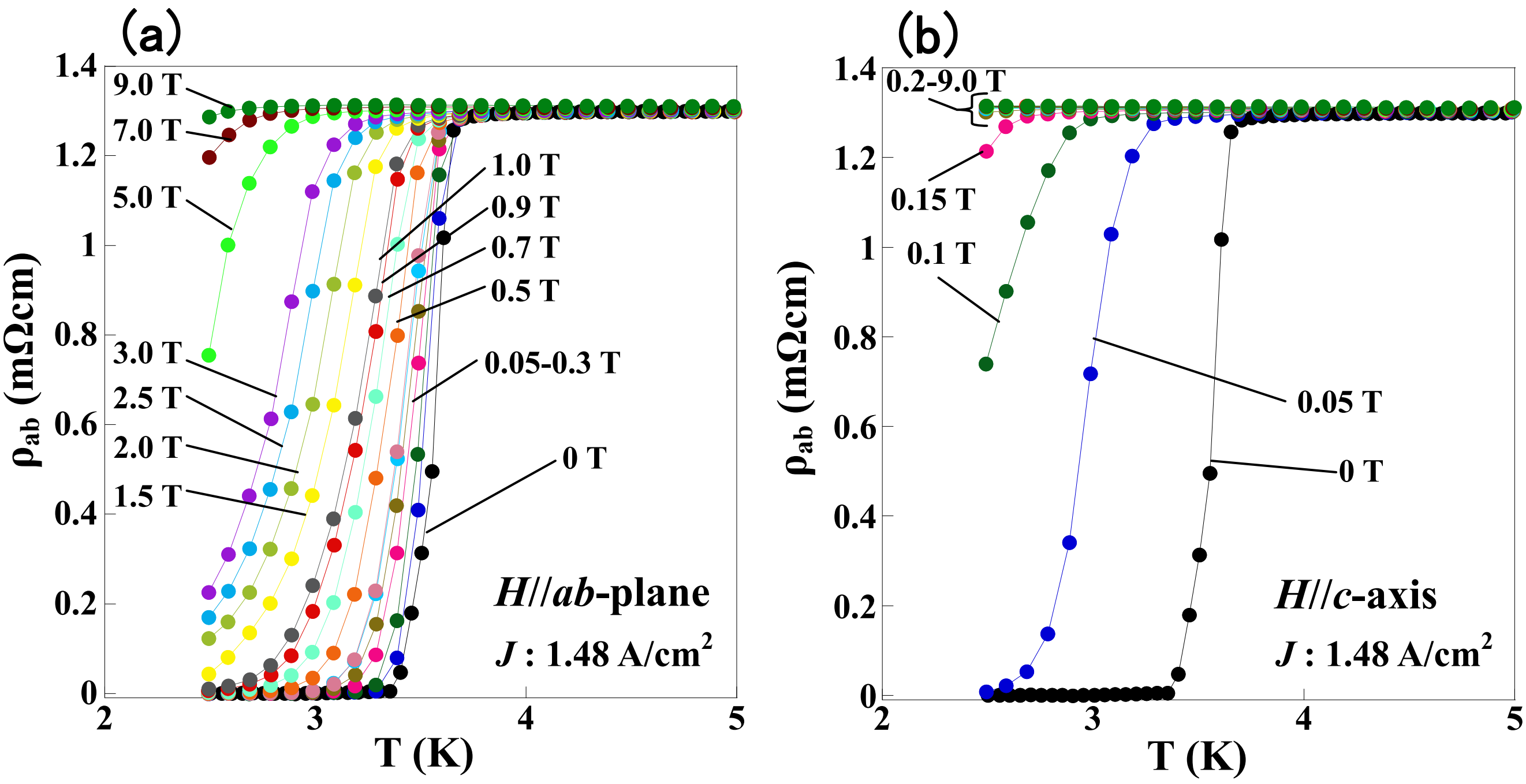

**Figure 4 (Color online)**

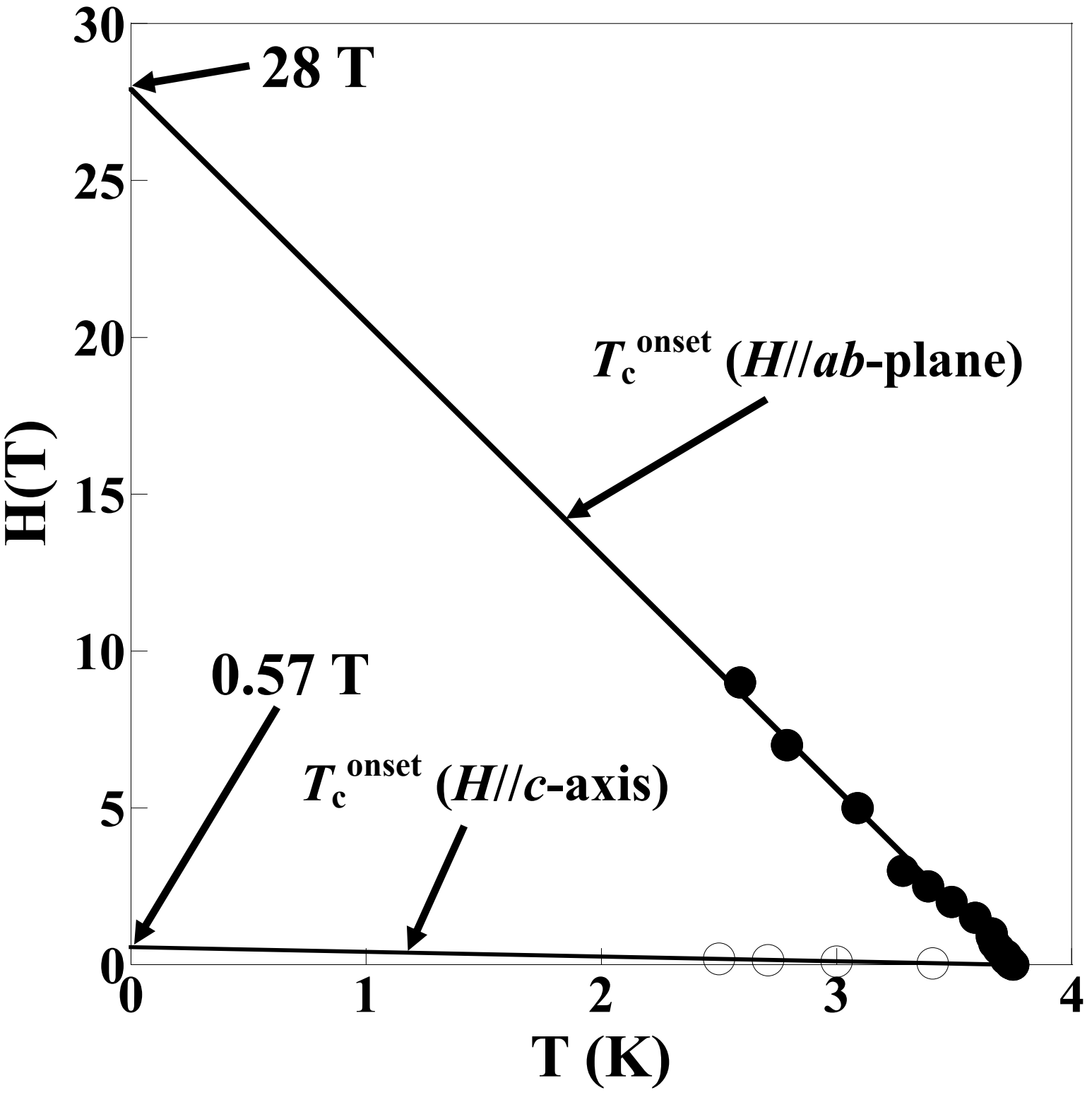

**Figure 5**

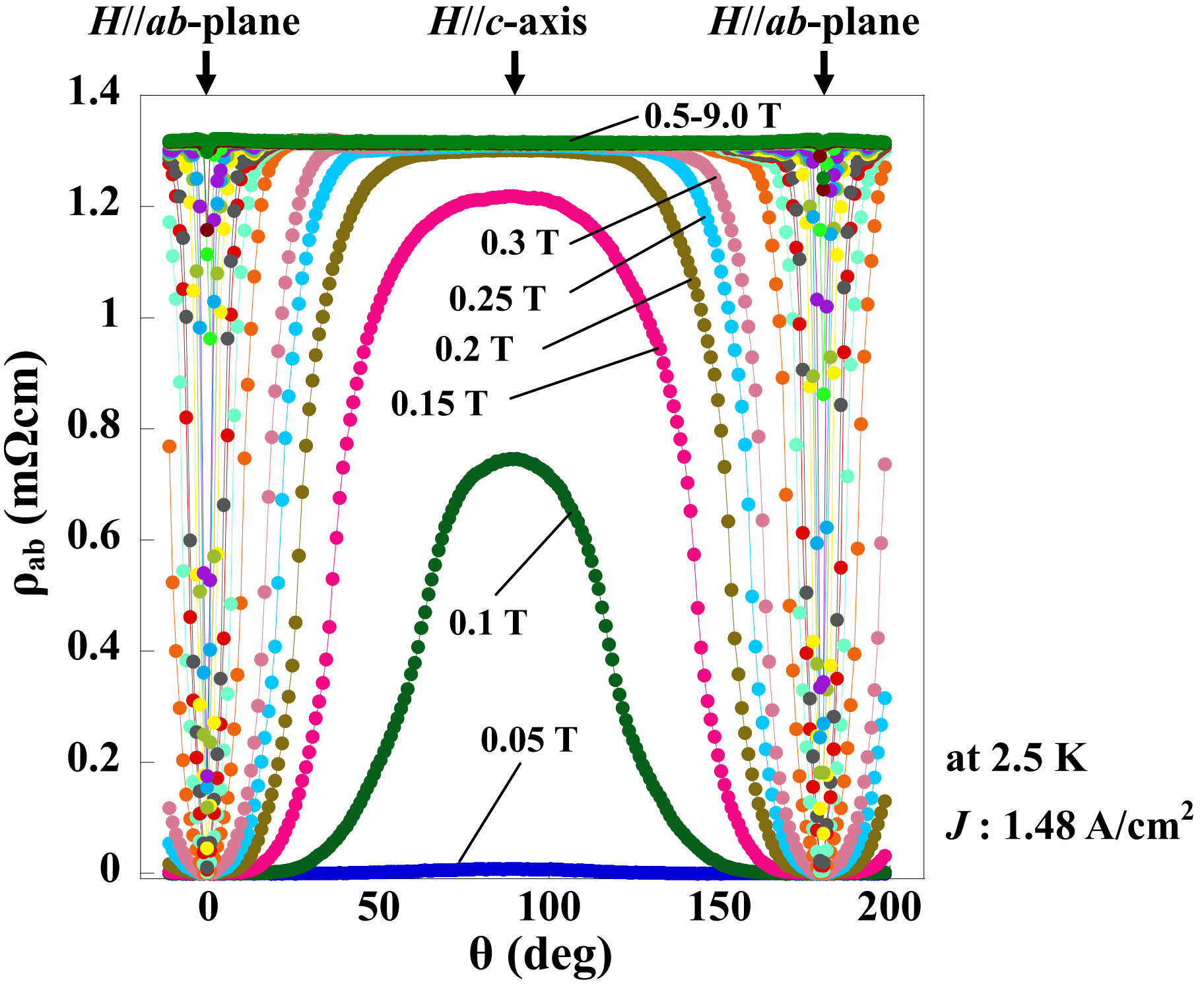

**Figure 6 (Color online)**

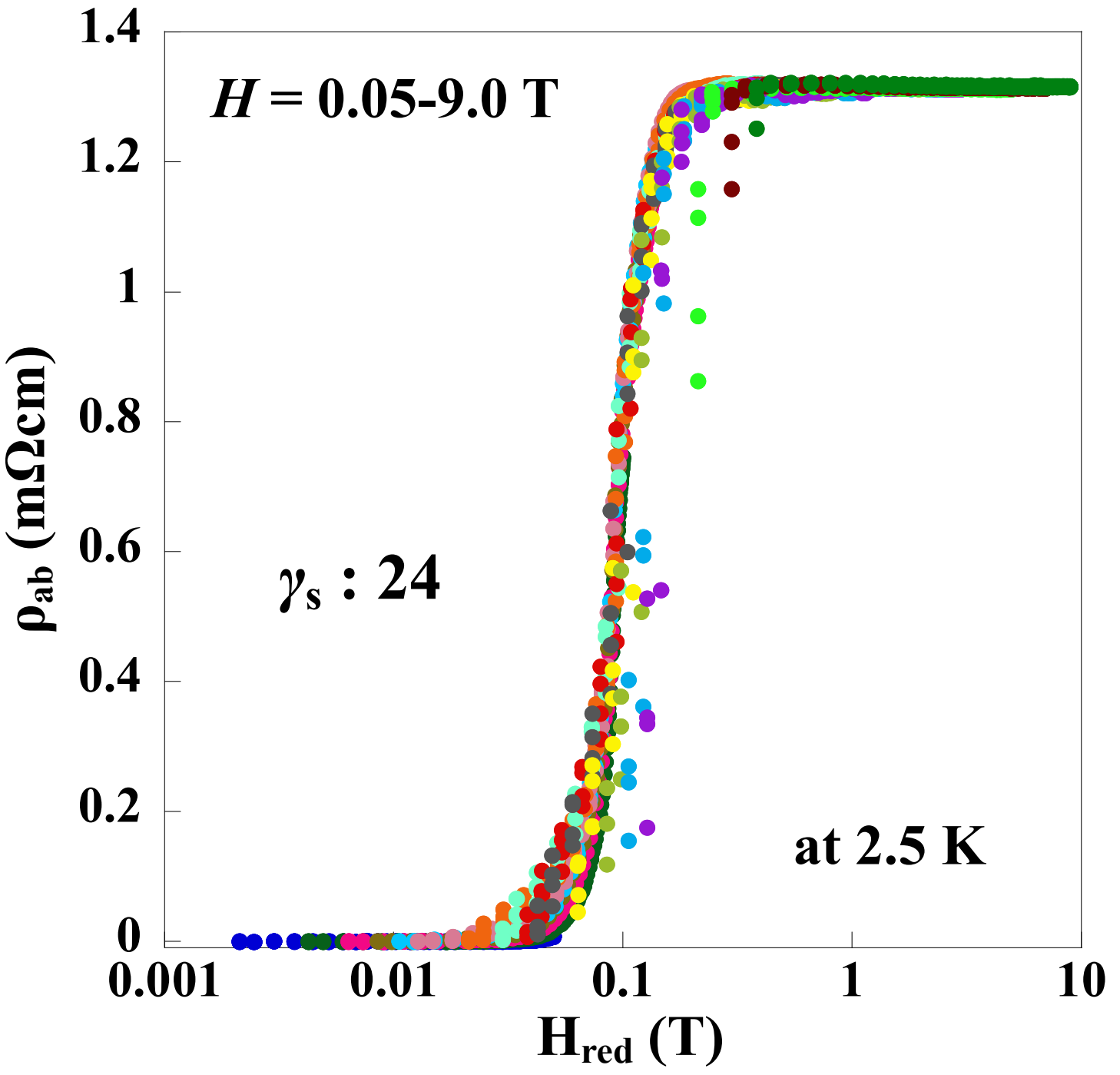

**Figure 7 (Color online)**